\documentstyle[11pt,conf_iap,epsf]{article}
\begin{document}
\heading{RR LYRAE STARS IN THE MACHO DATABASE} 

\photo{ }

\author{D. MINNITI$^{1}$,  C. ALCOCK$^{1,2}$, D.R. ALVES$^{1,4}$,  
R.A. ALLSMAN$^{2}$, T.S. AXELROD$^{1,10}$, A. BECKER$^{5}$,
D.P.  BENNETT$^{1,2}$, K.H. COOK$^{1,2}$, K.C. FREEMAN$^{10}$, 
K. GRIEST$^{2,6}$, J.A. GUERN$^{2,6}$, M.J. LEHNER$^{2,6}$,
S.L. MARSHALL$^{1,2}$, B.A. PETERSON$^{10}$, M.R. PRATT$^{1,5}$, 
P.J. QUINN$^{7}$, A.W. RODGERS$^{10}$, C.W. STUBBS$^{5}$, 
W. SUTHERLAND$^{8}$, D.L. WELCH$^{9}$ \\
(Le Groupe MACHO)}
       {$^{1}$ Lawrence Livermore National Laboratory, Livermore, CA 94550\\
       $^{2}$ Center for Particle Astrophysics, University of California, Berkeley, CA 94720\\
       $^{3}$ Supercomputing Facility, Australian National Univ., Canberra, ACT 0200, Australia\\
       $^{4}$ Department of Physics, University of California, Davis, CA 95616\\
       $^{5}$ Department of Physics and Astronomy, University of Washington, Seattle, WA 98195\\
       $^{6}$ Department of Physics, University of California San Diego, La Jolla, CA 92093-0350\\
       $^{7}$ European Southern Observatory, D-85748 Garching bei M\"unchen, germany\\
       $^{8}$ Department of Physics, University of Oxford, Oxford OX1 3RH, U.K.\\
       $^{9}$ Dept. of Physics and Astronomy, McMaster Univ., Hamilton, Ontario, Canada L8S 4M1\\
       $^{10}$ Mount Stromlo and Siding Springs Obs., Australian Natl. Univ., Weston, ACT 2611, Australia}

\bigskip

\begin{abstract}{\baselineskip 0.4cm 
The MACHO Project has catalogued $\sim 8000$ RR Lyrae stars in
the Large Magellanic Cloud, $\sim 1800$ in the Galactic bulge, and
$\sim$ 50 in the Sgr dwarf galaxy.
These variables are excellent distance indicators, and are used as tools
to study the structure of the Large Magellanic Cloud and the bulge.
The large datasets also probe uncommon pulsation modes.
A number of double-mode RR Lyrae stars (RRd) are found in
the Large Magellanic Cloud sample.  These stars provide important clues 
for understanding the formation and evolution of the
inner Galaxy, the Large Magellanic Cloud and the Sgr dwarf galaxy.
A large number of second overtone pulsators (RRe) are found in the LMC
and bulge. Finally, the RR Lyrae belonging to the Sgr dwarf 
yield an accurate distance to this galaxy. Their presence also alerts us of the
very interesting possibility of distant sources for bulge microlensing events.
}
\end{abstract}

\section{Introduction}

Even though all pulsating variables are interesting probes of the internal
constitution and evolution of stars \cite{GAU},
a subset of them, the RR Lyrae stars, are particularly good for astronomers 
for a variety of other reasons.

We use RR Lyr to measure distances within the Local Group.  
Having a narrow range of colors and luminosities, they are easy to 
identify from their characteristic variability.   

We also use RR Lyr to date
the Universe, and to study the formation of the Milky Way (MW), 
since they are among the oldest stars known. 

The MACHO project discovered large numbers of RR 
Lyr in the MW, the Large Magellanic Cloud (LMC), and the Sgr
dwarf galaxy (Sgr). Here we will discuss a few astronomical applications
of our on-going study of this sample, starting with a brief description of the
selection of RR Lyr among various other periodic variable stars (Section 2).

RR Lyr are excellent tools to study pulsational physics, and therefore
to determine fundamental stellar parameters. They can be used
to determine masses, the distance to the LMC, and the ages of the
oldest stellar populations, and to compare the old stellar 
populations of the LMC and the MW bulge (Section 3).
RR Lyr are primary distance indicators. They can be used to study the
structure of the inner MW (Section 4), and to determine the distance
and dynamical evolution of Sgr, thereby aiding in the interpretation of
the microlensing optical depth (Section 5).

\section{Selection of the Sample}

\begin{figure}
\begin{center}
\begin{minipage}{8cm}
\epsfxsize=12cm
\epsfbox{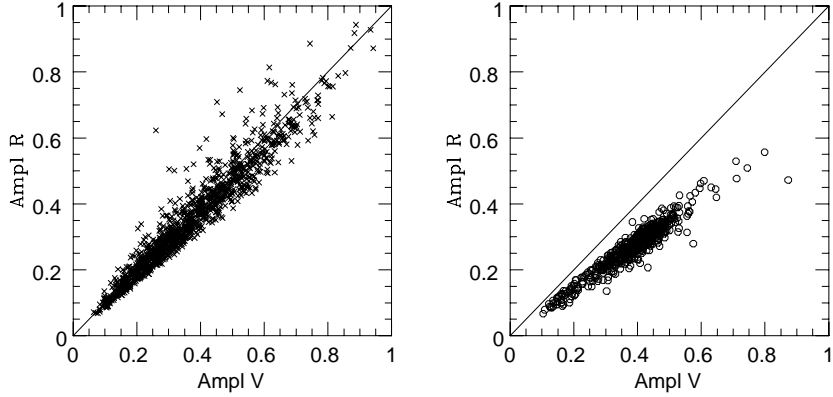}
\end{minipage}
\caption{
Red $vs$ blue amplitudes for all the variable stars with
$0.25^d < P < 0.4^d$ in the MACHO database. For binary stars $A_V
\approx A_R$. Note the clear separation from the RRc, which follow
the relation $A_R \approx 0.7 A_V$. This is very useful in order to
discriminate different variables, and it is possible in the MACHO
data because we have excellent coverage of the light curves in both
passbands.
}\label{iapf1}
\end{center}
\end{figure}

The MACHO database contains so far about 40000 periodic variable stars in
the LMC and a similar number in the MW bulge. A description of the 
variable star database has been published elsewhere \cite{COO}. Among these
stars, we have selected about 8000 RR Lyr in the LMC, about 1800 RRLyr in the 
MW bulge, and about 50 RR Lyr in Sgr. The selection process in the LMC
is straightforward \cite{AL1}, and here we will briefly describe the
selection of the bulge sample. The bulge RR Lyr were 
selected using the period-amplitude and amplitude-amplitude diagrams.
Here we analyze the data from 1993, covering over 100 days.  The selection 
of RR Lyrae in this database is complicated by aliases. Some of the stars with
periods $P = 1/n ~days$ maybe badly phased because the MACHO 
observations are done roughly every 24 hours. However, almost all 
aliases are discarded by demanding that $P_V=P_R$, which stresses the
importance of having continuous coverage in two passbands. 
The selection of RR Lyrae is also complicated by the presence of other 
variable stars with overlapping periods: eclipsing binaries, $\delta$ Scu stars,
etc.  Fortunately again, we have good quality light curves in two passbands.
The amplitude ratios can be used to discriminate the pulsating variable
stars from the eclipsing variables (Figure~\ref{iapf1}).

%

\begin{figure}
\begin{center}
\begin{minipage}{12cm}
\epsfxsize=12cm
\epsfysize=10cm
\epsfbox{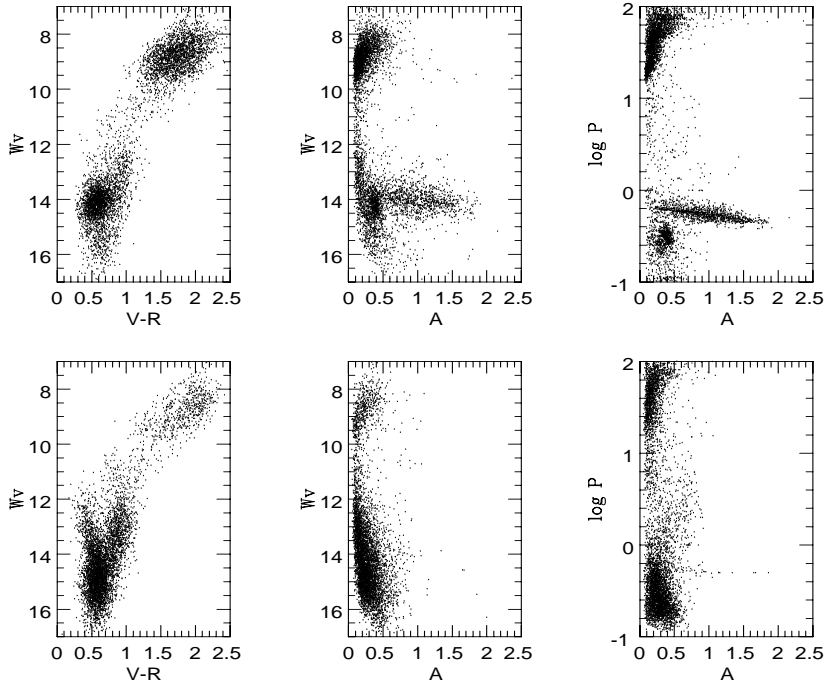}
\end{minipage}
\caption{
Separation between eclipsing (lower panels) and pulsating (upper
panels) variable stars with $0.1 < P < 100$ days in the MACHO 93 data.
From left to right we show the color-magnitude diagram (using the
reddening independent magnitudes $W_V$), the amplitude-magnitude diagram,
and the amplitude-period diagram.  In order of increasing periods, there are
contact and detached binaries, and spotted or RS~CVn-type stars in the
lower panels. The upper panels show in order of increasing periods
$\delta$~Scu, RRe, RRc, RRab, pulsating yellow variable,
and long-period variable stars. 
}\label{iapf2}
\end{center}
\end{figure}

The final RR Lyr sample is then selected from the period-amplitude
diagram, including all stars with $0.2^d < P < 1.1^d$, and $0.1 < A_V < 2$
(Figure~\ref{iapf2}).
The light curves of these objects were inspected to identify a few
surviving binary stars. As a final check, the light curves are fitted
by a series of Fourier sine functions \cite{SMI}. 
The resulting Fourier coefficients 
(e.g. the $R_{31}$ vs $\phi_{31}$ or the $\phi_{21}$ vs $\phi_{31}$ planes)
were used to discriminate among RR Lyr pulsating in the fundamental mode
(RRab), in the first-overtone (RRc), or in the second-overtone (RRe).   

This final sample is representative of the whole RR Lyr population in the 
bulge, but it is not complete.
Independent estimates of the completeness of our RR Lyr sample are obtained
by comparison with other surveys (\cite{ALA}, \cite{UD1}, \cite{BLA}).
Note that the number of bonafide RR Lyr in the MW bulge sample will increase
as the data from the following years is analyzed.

The photometric calibration of such a large database using non-standard
filters is challenging. We have used the latest calibrations 
\cite{ALV}, and have made a series of $external$ comparisons with other
photometry \cite{WA1}, \cite{UD1}, \cite{UD2}, \cite {CO2}.

%
%

Some of the MACHO bulge fields have significant overlap. 
The $internal$ comparisons of 44 bulge RR Lyr in overlap regions yield the 
following errors in the magnitudes $\sigma {V_M} = 0.120$, 
$\sigma {R_M} = 0.114$, in the positions $\sigma_{\alpha} = 0.58$'', 
$\sigma_{\delta} = 0.37$'', in the periods $\sigma_P = 0.000054^d$, and 
in the amplitudes $\sigma_{A V_M} = 0.12$, $\sigma_{A R_M} = 0.06$.

\section{Comparison Between RR Lyrae in the LMC and in the Bulge}

The RR Lyr in the MW bulge have mean $[Fe/H] = -1.1$, based on spectroscopic
determinations of 54 RR Lyr in Baade's window \cite{WA1}. 
This is significantly more metal-poor than the bulk of the bulge population,
which have mean $[Fe/H] = -0.2$ to $-0.6$,
as measured from spectroscopy of K giants in different bulge fields
\cite{MCW}, \cite{MI2}.
This is not surprising, however, since we know from globular clusters that
metal-rich populations do not produce as many RR Lyr \cite{SUN}, \cite{LAY}.

\begin{figure}
\begin{center}
\begin{minipage}{8cm}
\epsfxsize=12cm
\epsfbox{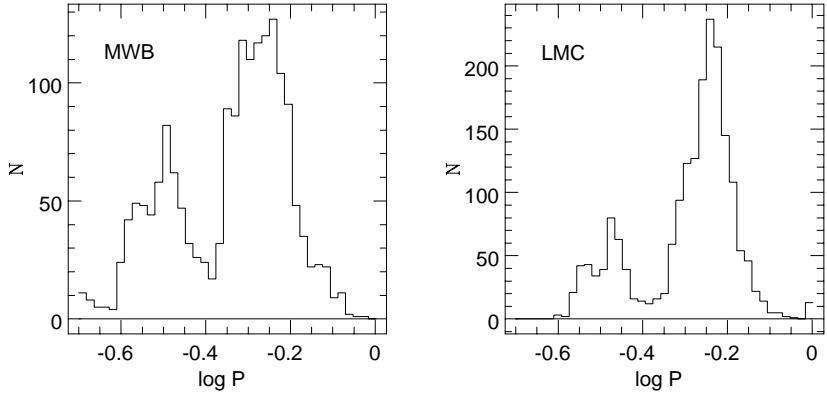}
\end{minipage}
\caption{
Period distribution for RR Lyr stars in the LMC (right panel),
and the Galactic bulge (left panel). The peaks correspond, from right
to left, to RRab, RRc, and RRe, respectively. Note that the overall
distribution of bulge RR Lyr is shifted towards shorter periods.
This is a metallicity effect, with the bulge RRab being about $0.5~ dex$
more metal-rich in the mean than the RRab in the LMC.
}\label{iapf3}
\end{center}
\end{figure}

%


\begin{center} 
{\bf Table 1.} Mean Periods for LMC and Bulge RR Lyr
\end{center}
\begin{center}
\begin{tabular}{|l|l|l|l|l|}
\hline 
Type & Puls. Mode & $P_{LMC}$ & $P_{MWB}$ & $WV_{MWB}$ \\
\hline
     &        &      &     & \\
RRab & Fundamental & 0.55$^d$ & 0.53$^d$ &14.05\\
RRc  & 1st Overtone& 0.34$^d$ & 0.32$^d$ &14.29\\
RRe  & 2nd Overtone& 0.28$^d$ & 0.27$^d$ &14.39\\
RRd  & Double & 0.48$^d$ &      &     \\
\hline
\end{tabular}
\end{center}
\medskip

\begin{figure}
\begin{center}
\begin{minipage}{8cm}
\epsfxsize=12cm
\epsfbox{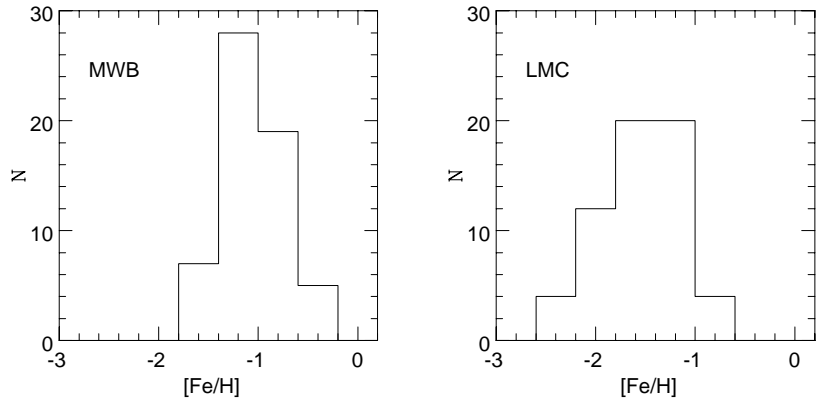}
\end{minipage}
\caption{
Metallicity distribution of RR Lyr in Baade's Window from Walker \& Terndrup (1992),
and for the LMC from Alcock et al. (1995). The histograms have been arbitrarily
normalized to the same total number of counts.
}\label{iapf4}
\end{center}
\end{figure}

Alcock et al. \cite{AL1} studied the RR Lyr in the LMC, here we compare that 
sample with the RR Lyr in the MW bulge.
The RR Lyr in the LMC and in the bulge have very different 
metallicity distributions (Figure~\ref{iapf4}). Their mean metallicities differ by 
$\Delta [Fe/H] = 0.5 ~dex$, with the LMC ones being more metal-poor. 
This has to be one of the major parameters
driving the different distribution of RR Lyr in the Bailey diagram \cite{BO1},
and in their different period distribution,
as shown in Figure~\ref{iapf3}.

The transition periods (defined based on the longest period RRc star)
in the bulge and LMC are $P_{tr} = 0.432^d$, and $P_{tr} =
0.457^d$, respectively. The RR Lyr population in the LMC is Oosterhoff type I, 
while in the bulge it is extreme Oosterhoff type I.

In particular,
the mean periods of the RR Lyr pulsating in the different modes are longer
in the LMC than in the MW bulge, as listed in Table 1. Also shown in Table
1 are the mean reddening independent magnitudes $W_V$ for bulge RR Lyr 
pulsating in the fundamental, 1st overtone, and 2nd overtone, respectively.

%

\begin{figure}
\begin{center}
\begin{minipage}{10cm}
\epsfxsize=10cm
\epsfxsize=8cm
\epsfbox{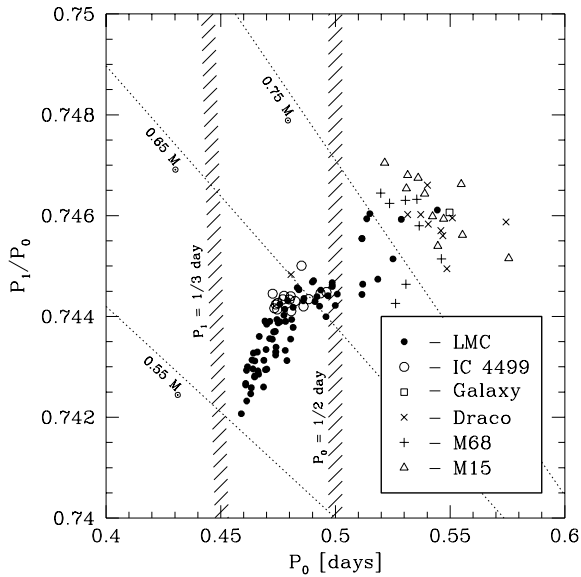}
\end{minipage}
\caption{
Petersen diagram for RRd in the LMC taken from Alcock et al. (1996b).
The ridge lines corresponding to
different masses from the pulsation equation are shown. A wide range of
masses is evident in the LMC RRd stars.
Note that both $P_1$ and $P_0 / P_1$ are very accurately measured.
We strongly caution that the whole
parameter space has not been fully exploited. In particular, there may be
more RRd with shorter and longer periods that those shown here.
The lower limit to the masses will be dictated by bare stellar cores, with
$M \geq M_{WD}$. Conversely, because of mass loss during the RGB phase,
the upper limit to the masses is constrained by the mass of
turn off stars, $M \leq M_{TO}$.
}
\label{iapf5}
\end{center}
\end{figure}

%
%

The large sample sizes allow us to identify uncommon pulsation modes,
such as double-mode RR Lyr or RRd. While the bulge database has not been 
searched yet, we have identified more than 70 RRd in the LMC \cite{AL3}. These
are double-mode pulsators, with typical light curves as shown in \cite{AL1}.
We use the Petersen diagram \cite{PET} to measure the masses of these stars
(Figure~\ref{iapf5}). 
Knowing the masses, we can obtain their intrinsic luminosities from the
pulsation equation \cite{BO3}, which in turn provides with a measure of the
distance to the LMC. We can then solve the old problem of the different LMC 
distance scales produced by the Cepheids and RR Lyrae \cite{WA1}.  The results 
of this revised LMC distance scale will be published elsewhere \cite{AL3}.

The existence of RR Lyr pulsating in the 
second overtone or RRe have been difficult to prove \cite{SMI}, \cite{SMO}. 
Large numbers of RRe have been now identified in the LMC and the MW bulge. 
In the mean, bulge RRe candidate stars are
about 0.1 mag fainter than RRc stars (Table 1). These are not to be confused
with $\delta$~Scu stars in the bulge, which are more than a magnitude fainter
in the mean. Further study will reveal
more fundamental properties of these candidate RRe stars.

%
%
%
%
%

\section{Spatial Distribution of Bulge RR Lyrae}

\begin{figure}
\begin{center}
\begin{minipage}{8cm}
\epsfxsize=8cm
\epsfbox{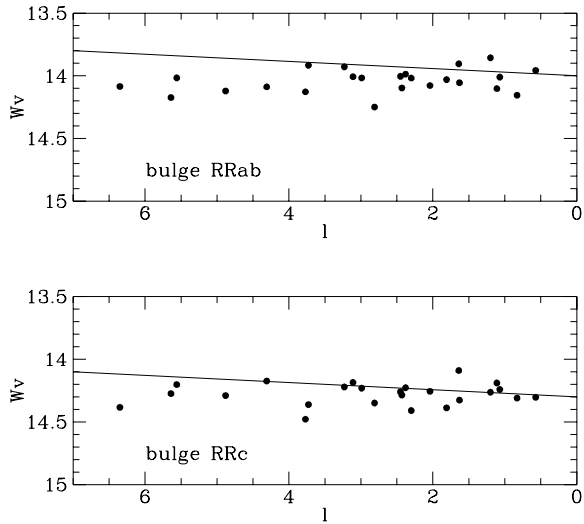}
\end{minipage}
\caption{
Mean magnitudes of RRab (top panel) and RRc (bottom panel) for
the 24 MACHO bulge fields as function of Galactic longitude.
The solid line shows the trend expected from a barred distribution.
This figure shows that the RR Lyr do not follow the barred distribution
seen in the clump giants and other tracers.
}\label{iapf7}
\end{center}
\end{figure}

There is a bar in the inner MW, seen from the integrated IR light
\cite{BLI}, from tracers of metal-rich populations such as Miras
\cite{WHI}, and from the kinematics of gas \cite{BIN}, and stars \cite{ZHA}.
This bar is well described by the density distribution 
$\rho_{b} \propto r^2~exp( 1/2 [(((r~cos~b~cos l - R_0) / 1.49)^2 +
(r~cos b~sin l / 0.58)^2)^2 + (r~sin b / 0.40)^4]^{1/2}$ modeled from
the IR maps of COBE \cite{DWE}.
Clump giants have also been used to study the spatial distribution of the MW
bulge. The clump giants show a barred distribution seen in 
other sources, with the closer side of the bar located at positive longitudes
\cite{ST1}, \cite{ST2}. 
There is a clear magnitude difference between the near and far sides of the
bar, about 0.5 mag every 20 degrees.   The RR Lyr are
excellent distance indicators, and should show the bar effect if this
is present. We have computed the mean magnitudes of the bulge RR Lyr 
in the different MACHO fields. 
Figure~\ref{iapf7} shows the mean reddening-independent magnitudes of the
RRab and RRc in each of the MACHO bulge fields. Each point comes
from averaging 20-90 stars. RR Lyr belonging to the foreground or to the Sgr
dwarf have been discarded before computing these mean magnitudes.
It is evident that the RR Lyr do not follow the expected
barred distribution indicated by the solid line. This result confirms 
previously unpublished results from the Palomar-Groningen field \cite{WES}.

The morphology of the Bailey diagram (amplitude-period diagram) is 
determined in part by the metallicity of the population \cite{BO1}.
This diagram allows us to obtain relative abundances, and to divide the 
RRab sample into three bins, containing metal-poor, intermediate metallicity, 
and metal-rich RRab stars.
We have computed the mean $W_V$ magnitudes of these
three RRab subsamples of different metallicities.
These do not show a barred distribution, with the
possible exception of the most metal-rich ones.
For the metal-rich subsample, the bar cannot be ruled out within the errors.
Note that a potentially more precise way to measure metallicities directly from 
their light curves requires the use of the Fourier coefficients \cite{KOV}.

Having established that the line of sight distribution of RR Lyr is not
barred, we will now
examine the radial dependence of their surface density distribution.
This is complicated by effects such as incompleteness, the line of sight 
depth of the bulge,  the presence of a 
mixed population and of a metallicity gradient. On top of these, reddening
is always a concern in bulge fields \cite{NG1}. We used reddening 
independent magnitudes, defined as $W = V - 3.97~(V-R)$.
In order to evaluate the completeness, we will use the counts of bright W UMa
systems in the bulge.  Because of our magnitude limits,
these systems belong to the disk, and their distribution
should be uniform in the MACHO fields. W UMas are very numerous
variable stars with similar
periods and amplitudes as the bulge RR Lyr, and therefore provide an
excellent check on our efficiency to find RR Lyr in the different bulge
fields. In the innermost bulge fields, where crowding is severe and 
incompleteness may be higher than $50\%$, the ratio
of relative numbers $N_{RR} / N_{WUMa}$ would give a better indication of the
real density profile of the RR Lyr population. The price to pay here is
that we add the Poisson noise of $N_{WUMa}$. While the relative count
normalizations in different fields are obtained with the help of the W UMas,
the absolute normalization is uncertain. We use the total number
of MACHO RR Lyr discovered in Baade's Window, which is the most
thoroughly studied bulge field, complemented
with the searches of \cite{BLA}, \cite{UD1}, and \cite{UD2}.

In these inner MW fields we have to consider the contribution from
all possible Galactic components, namely disk, bulge and halo. They
have different radial density profiles, and we can try to relate the RR Lyr
density distribution to one of these components.

The MW disk behaves as a double exponential, 
$\rho \propto e^{-r/h_r} e^{-z/h_z}$, 
with scalelength $h_r = 3.5$ kpc, and scaleheight $h_Z = 0.3$ kpc \cite{OJH}.
The peaked magnitude distribution of the RR Lyr in the MACHO fields 
(Figure 1 of \cite{AL2}), as well as their concentrated spatial distribution,
rules out a significant contribution from disk RR Lyr. 
The W UMas, however, are useful tracers of the disk, as noted above. We will
use these variables to study the structure of the inner disk in a future paper,
following \cite{RU3}.

The MW bulge shows a very steep density distribution along the bulge
minor axis, traced by
K and M giants, IRAS sources, and integrated K-band light. This radial density
distribution is represented by 
a power law, $\rho_b \propto r^{-m}$, with $m=3.7-4.2$ \cite{FRO}.
Such a steep power law with $m=3.7-4.2$ does not fit our observations.

The MW halo outside of the bulge region ($R>3~ kpc$), traced by
halo globulars \cite{ZIN}, RR Lyrae stars \cite{SAH}, field BHB stars 
\cite{PRE}, also shows a power law density distribution, 
$\rho_h \propto r^{-n}$, although shallower, with $n=3.0-3.5$.
The RR Lyr in the MACHO fields are more consistent with the
extrapolation of such a power law with $n=3.0-3.2$ for $0.3<R<1 ~kpc$,
extending the results on the Palomar-Groningen field at $R = 1$ kpc
\cite{WES}.

Once again, the radial density profile of bulge RR Lyr is different from the
bulk of the metal-rich bulge component represented by K and M giants.
The most straightforward interpretation is that these represent different 
populations, tracing different components of the MW. The RR Lyr would belong 
to the inner extension of the halo, which is relatively metal-poor, 
while the dominant metal-rich component
traced by the clump giants would represent the bar \cite{MI1}.

\section{RR Lyrae in the Sgr Dwarf}

\begin{figure}
\begin{center}
\begin{minipage}{8cm}
\epsfxsize=12cm
\epsfbox{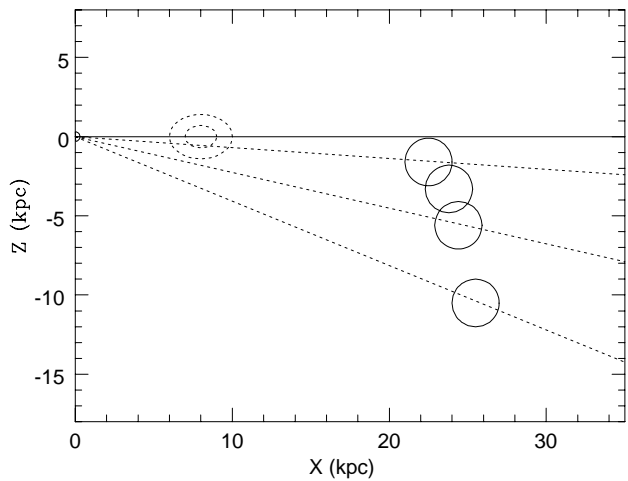}
\end{minipage}
\caption{
Galactic {\it meridien de Paris}, showing
schematically the distances to the Sgr dwarf
galaxy measured at different Galactic latitudes using RR Lyr stars
(large circles). The sizes of the circles roughly scale with the errors
in the distance determinations. From top to bottom, the MACHO distance,
the DUO distance, and two OGLE distances.  Paris is at (0,0), and the
MW bulge is at (8,0).
}\label{iapf8}
\end{center}
\end{figure}

From the magnitude distribution of RR Lyr in the MACHO bulge database we
found about 50 RR Lyr that are fainter than a typical bulge RR Lyr by
$\sim 2.2$ mag. These belong to the Sgr dwarf \cite{AL2}, extending even
more the known size of Sgr: its major axis would now cover $\geq 20^{\circ}$
in the sky.   These faint RR Lyr can be used
to obtain an accurate distance of Sgr in the MACHO fields.  This distance 
turns out to be smaller than that measured in other fields
\cite{MM1}, \cite{MM2}, \cite{ALA}, as shown in Figure~\ref{iapf8}. This has allowed
us to obtain the 3-dimensional shape of the Sgr dwarf.

%

The discovery of Sgr RR Lyr in the bulge fields is significant because 
the observed microlensing optical depth towards the bulge was larger than 
expected \cite{KIR}. Most, if not all, of the large optical depth may be 
explained by the presence of a bar in the inner MW, pointing almost towards 
the Sun \cite{PA1}.  However, 
MACHO has detected 100 microlensing events towards the MW bulge as of June 
1996. As the sample increases in size,
it is tempting to consider that a small fraction of the events
may have sources in the Sgr dwarf \cite{DVG}, \cite{AL2}.
Note that Sgr will provide a relatively small number of sources, but the
number of bulge lenses is large. Because we know the geometry, as well as
the relative velocities involved, this opens up the interesting possibility
of directly measuring the mass of a lens for the configuration
(source, lens) = (Sgr, bulge). A first, and relatively simple,
step towards identifying possible Sgr sources would be
to acquire accurate radial velocities for the observed events.

%
%
%
%
%
%

\section{Summary}

We are exploring the RR Lyr in the MACHO database. A comparison between the
properties of the RR Lyr in the LMC, the MW bulge, and the Sgr dwarf yields
several interesting results:

$\bullet$  The large number of RR Lyr in the LMC and the MW bulge allowed us
to identify uncommon pulsation modes.

$\bullet$  LMC and MW bulge RR Lyr have different properties, 
mostly due to their different metallicities.

$\bullet$ The RR Lyr population in the MW bulge is not barred, in contrast with
the clump giants.


$\bullet$ The Sgr dwarf is very extended, $> 20^{\circ}$ along its major axis.

$\bullet$ The number of microlensing events detected in the MW bulge is large
enough ($>100$) that the sample may already contain some due to
sources in the Sgr dwarf.

\acknowledgements{
We are very grateful for the skilled support by the technical staff at MSO.  
Work at LLNL is supported by DOE contract W7405-ENG-48. 
Work at the CfPA is supported NSF AST-8809616 and AST-9120005.  
Work at MSSSO is supported by the Australian Department of Industry, 
Technology and Regional Development.  
WJS is supported by a PPARC Advanced Fellowship.  
KG thanks support from DOE OJI, Sloan, and Cottrell awards.
CWS thanks support from the Sloan, Packard and Seaver Foundations.
}

\vfill
\end{document}